\documentclass[prl,aps,superscriptaddress,floatfix,tightenlines,twocolumn]{revtex4}

\usepackage{amsmath}
\usepackage{amssymb}
\usepackage{amsfonts}
\usepackage{epsfig}
\usepackage{subfigure}
\usepackage[dvips]{color}
\usepackage{graphicx}
\usepackage{bm}% bold math
\usepackage{times}
\usepackage{amsthm}

\begin{document}
\title{Demonstration of Entanglement-Enhanced Phase Estimation in Solid}
\author{Gang-Qin Liu}
\affiliation{Beijing National Laboratory for Condensed Matter Physics, Institute
of Physics, Chinese Academy of Sciences, Beijing 100190, China}
\author{Yu-Ran Zhang}
\affiliation{Beijing National Laboratory for Condensed Matter Physics, Institute
of Physics, Chinese Academy of Sciences, Beijing 100190, China}
\author{Yan-Chun Chang}
\affiliation{Beijing National Laboratory for Condensed Matter Physics, Institute
of Physics, Chinese Academy of Sciences, Beijing 100190, China}
\author{Jie-Dong Yue}
\affiliation{Beijing National Laboratory for Condensed Matter Physics, Institute
of Physics, Chinese Academy of Sciences, Beijing 100190, China}
%\author{...}
\author{Heng Fan}
\email{hfan@iphy.ac.cn}
\affiliation{Beijing National Laboratory for Condensed Matter Physics, Institute
of Physics, Chinese Academy of Sciences, Beijing 100190, China}
\affiliation{Collaborative Innovation Center of Quantum Matter, Beijing 100190, China}
\author{Xin-Yu Pan}
\email{xypan@aphy.iphy.ac.cn}
\affiliation{Beijing National Laboratory for Condensed Matter Physics, Institute
of Physics, Chinese Academy of Sciences, Beijing 100190, China}
\affiliation{Collaborative Innovation Center of Quantum Matter, Beijing 100190, China}
\date{\today}
%\pacs{ }
\begin{abstract}
Precise parameter estimation plays a central role in science and technology.
The statistical error in estimation can be decreased by repeating measurement,
leading to that the resultant uncertainty of the estimated
parameter is proportional to the square root of the number of repetitions in
accordance with the central limit theorem. Quantum parameter estimation, an
emerging field of quantum technology, aims to use quantum resources to yield
higher statistical precision than classical approaches. Here, we report the first
room-temperature implementation of entanglement-enhanced phase estimation
in a solid-state system: the nitrogen-vacancy centre in pure diamond.
We demonstrate a super-resolving phase measurement with two entangled qubits
of different physical realizations: an nitrogen-vacancy centre electron spin and a proximal
${}^{13}$C nuclear spin. The experimental data shows clearly the uncertainty
reduction when entanglement resource is used, confirming the theoretical expectation.
Our results represent an elemental demonstration of enhancement of quantum metrology against classical procedure.
\end{abstract}
\maketitle

Information about the world is acquired by observation and measurement, the results
of which are subject to error \cite{H}. The classical approach to reduce the statistical
error is to increase the number of resources for the measurement in accordance with
the central limit theorem, however, this method sometimes seems undesirable and inefficient
\cite{IRS}. Quantum parameter estimation, the emerging field of quantum technology, aims to
yield higher statistical precision of unknown parameters by harnessing entanglement and other
quantum resources than purely classical approaches \cite{review}. Since this
quantum-enhanced measurement will benefit all quantitative science and technology, it has
attracted a lot of attention as well as contention. Using $N$ independent particles to estimate
a parameter $\varphi$ can achieve at best the standard quantum limit (SQL) or called shot noise
limit scaling as $\delta\varphi\propto1/\sqrt{N}$ while it is believed that using $N$ entangled
particles and exotic states such as NOON states in principle is able to achieve the inviolable
Heisenberg limit scaling as $\delta\varphi\propto1/N$ \cite{Lloyd2006,zzb1}.  In such
circumstances, there are many efforts using non-classical states and quantum strategy for
sub-SQL phase estimation in different physical realizations, such as optical interferometry
\cite{IRS,opt1,opt2,opt3,opt4}, atomic systems \cite{atom1,atom2}, and Bose-Einstein condensates
\cite{BEC1,BEC2}.

In this paper, we report the first room-temperature proof-of-principle implementation of
entanglement-enhanced phase estimation in a solid-state system: the nitrogen-vacancy (NV)
centre in pure diamond single crystal. An individual NV center can be viewed as a basic unit of a quantum
computer in which the nuclear spin with a long coherence time performs as the memory and the
centre electron spin with a high control speed acts as the probe. This solid-state system is one
of the most promising candidates for quantum information processing (QIP), and many coherent
control and manipulation processes have been performed with this system
\cite{Gruber97Science,Fedor04PRLCNOT, Childress06Science, Lukin07ScienceRegister, Wrachtrup08entangle, SingleShotReadout,
Hanson11Nature, 1S, GatewhileDD, DuDJA, DDGate, DuCDD, ErrorCorr2014WangY, ErrorCorr2014Hanson, ourgroup1, ourgroup2}.
Here, we demonstrate a super-resolving phase measurement with two entangled qubits of different physical realizations:
a NV centre electron spin and a proximal ${}^{13}$C nuclear spin. We are able to improve the phase
sensitivity by factors close to $\sqrt{2}$ compared with the classical scheme, which conforms to the
fundamental Heisenberg limit. As we have entangled two qubits with different physical realizations,
our results represent a more generalized and elemental demonstration of enhancement of quantum
metrology. Moreover, our system has overcomed the defects of post-selection in the most common
optical systems which are fatal due to the fact that the measurement trials abandoned will eliminate
the quantum advantage over classical strategy.

\section*{Results}

\subsection{System description}
\begin{figure}[b]
\begin{center}
 \includegraphics[width=7.8cm]{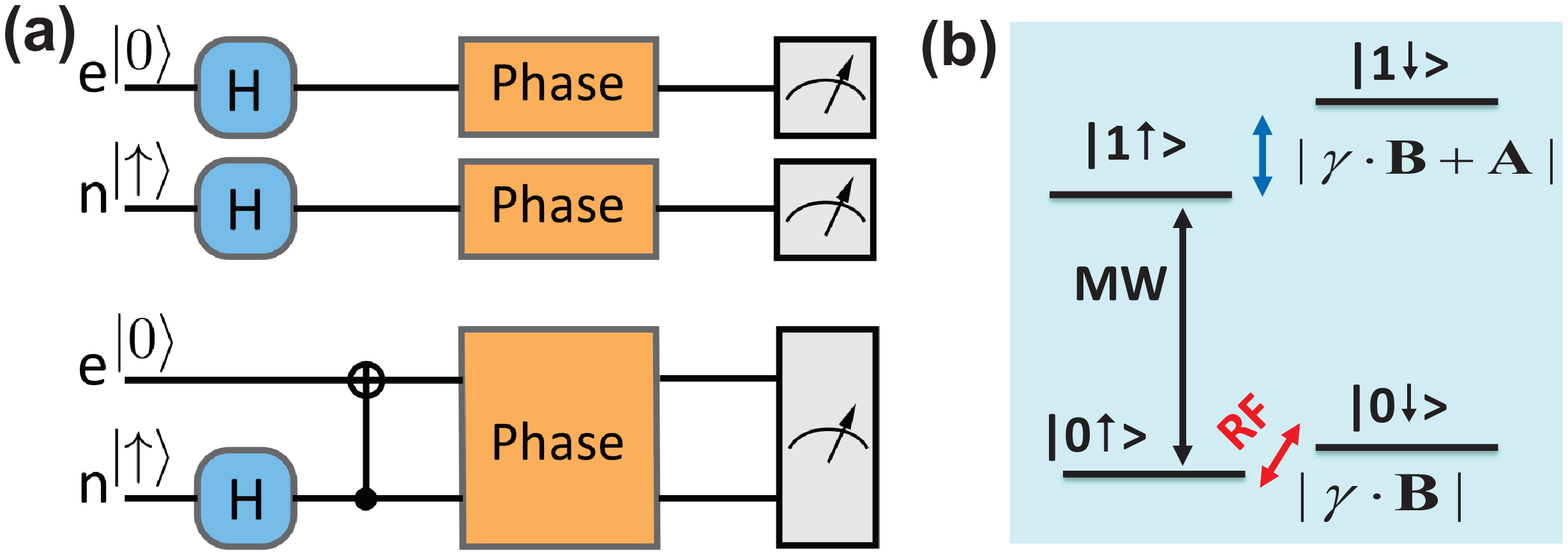}
\caption{\textbf{General scheme and system description.} \textbf{(a)} Phase estimation schemes of the independent
states and the electron-nuclear entangled state. {By harnessing entanglement, quantum metrology yields higher statistical precision than classical approaches. \textbf{(b)} Energy levels and physical encoding of the two-bit system. The electron spin and a nearby $^{13}$C nuclear spin of an NV center are employed to demonstrate the metrology scheme. At excited-state level anti-crossing (ESLAC), both spins can be polarized, manipulated and readout with high fidelity. Two-bit conditional quantum gates are implemented by applying selective microwave (MW) or (RF) pulses.}}
\end{center}
\label{Fig 1}
\end{figure}
 The phase estimation scheme is implemented by optically detected magnetic resonance (ODMR)~\cite{Gruber97Science, Childress06Science}
 technique on a home-built confocal microscope system. The description of the system can be found in \cite{DDGate}.
 The spin-1 electron spin of NV centre has triplet ground states with a zero-field splitting of $\Delta\approx
2.87$~GHz between the states $|0\rangle$ and $|\pm 1\rangle$. As an external magnetic field of about {507}~Gauss
 is applied along [111] direction of the diamond crystal, the degeneration of $|\pm 1\rangle$ states can be well relieved,
  and the first qubit is encoded on the $|0\rangle$ and $|-1\rangle$ subspace ((hereinafter labelled as $|1\rangle$)).
{ The electron spin state can be initialized to $|0\rangle$ state by a short 532~nm laser pulse (3~$\mu $s)
  and manipulated by resonant microwave (MW) pulses of tunable duration and phase.}
  The electron spin state is readout by collecting the spin-dependent fluorescence. To enhance the fluorescence collection efficiency, a solid immersion lens (SIL) \cite{SIL} is etched above the selected NV center, typical count rate in this experiment is 250~kps with SIL{, see Methods for details}.

The second qubit is encoded on the $|\uparrow\rangle$ and $|\downarrow\rangle$ states of a nearby $^{13}$C nuclear spin.
{See Fig.~1(b) for the energy levels of the two qubit system.}
The coupling strength between the target nuclear spin and centre electron spin is 12.8~MHz,
which indicates the $^{13}$C atom sites on the third shell from the NV centre \cite{Hyperfine}.
The polarization and readout procedure of the nuclear spin is more complicated than that of a electron spin.
The {507}~Gauss magnetic field causes excited-state level anti-crossing (ESLAC) of centre electron spin,
in which the optical spin polarization of centre electron will transfer to nearby nuclear spins \cite{ESLAC09PRL, ESLAC09PRA}.
So the host $^{14}$N nuclear spin, the nearby $^{13}$C nuclear spin as well as the center electron spin
 are polarized by the same laser pulse under this magnetic field.
To readout the nuclear spin state, a mapping gate, which transfers nuclear spin state to electron spin,
and a following optical readout of electron spin state are employed \cite{Lukin07ScienceRegister, ESLAC09PRA}, {see Methods for details}.
%The state manipulation of nuclear spin is similar to electron spin's, but with much lower frequency(in Radio frequency( RF) region).

\begin{figure}[t]
\begin{center}
\includegraphics[width=8.5cm]{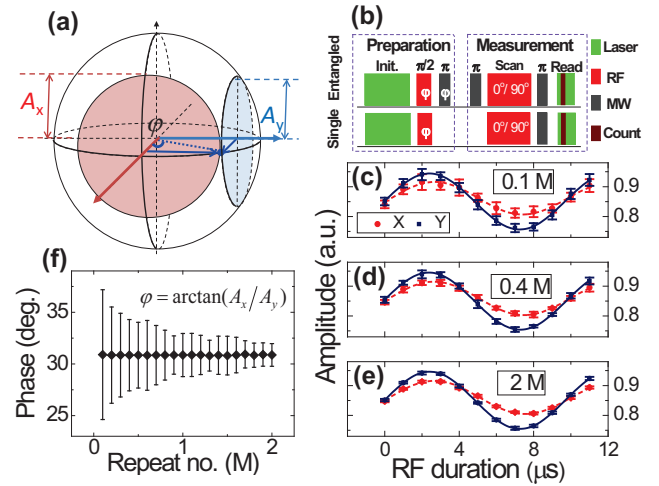}
\caption{\textbf{Phase preparation and measurement} \textbf{(a)} For each unknown state sitting on the equatorial plane of the Bloch sphere,
 {we compare the amplitudes of the two Rabi signals from this state (under orthogonal microwave pulses driven) to extract the original phase information:
 $\varphi=\arctan(A_x/A_y)$. %, where $A_x$ and $A_y$ are the amplitudes of Rabi signals
  %driven by $0^\circ$ and $90^\circ$ MW (or RF) pulses, respectively.
  \textbf{(b)} Upper pane: pulse sequence to prepare and measure electron-nuclear spin entanglement state $\frac{1}{\sqrt{2}}(|1\uparrow\rangle+e^{2i\varphi}|0\downarrow\rangle)$. Lower pane: pulse sequence to prepare and measure nuclear spin superposition state $\frac{1}{\sqrt{2}}(|0\uparrow\rangle+e^{i\varphi}|0\downarrow\rangle)$.
  \textbf{(c-e)} Nuclear Rabi signals for phase measurement, with 0.1~M, 0.4~M and 2~M repetition of pulse sequence in \textbf{(b)}. The input phase $\varphi$ is $30 ^\circ$. Solid circle with blue fitting line is driven by $0^\circ$ RF pulse (X measurement), and square with red dash fitting line is driven by $90^\circ$ RF pulse (Y measurement).  \textbf{(f)} Dependence of measured phase and its standard deviation on repeat number.}}
\end{center}
\label{Fig_2}
\end{figure}

The nearby nuclear spin couples to the centre electron spin through
strong dipolar interaction, which provides excellent conditions to implement two-qubit controlled gate.
 On the one hand, the resonant frequency of $|0\uparrow\rangle \Leftrightarrow|1\uparrow\rangle$ transition and
  $|0\downarrow\rangle \Leftrightarrow|1\downarrow\rangle$ transition are separated by 12.8~MHz from each other,
  so we can selectively manipulation one branch of nuclear spin with high fidelity while keep the other branch untouched {(using weak MW pulses, see black arrow in Fig.~1(b))}.
  On the other hand, the nuclear spin state evolution is strong affected by the state of electron spin:
   when electron spin is on the $|0\rangle$ state (or $|-1\rangle$ state), the dynamics of the nuclear spin is dominated by the
  external magnetic filed (or the dipolar interaction, respectively),
   its Zeeman splitting between the $|\uparrow\rangle$ and $|\downarrow\rangle$ states
is about {500~kHz}, which is far away from the dipolar interaction strength of 12.8~MHz.
Therefore, we can selectively manipulate nuclear spin state in one branch of electron spin, as well
{(using RF pulses, see red arrow in Fig.~1(b)  and Supplementary Note 1)}.

\subsection{Phase {preparation and} measurement}

{Fig.~2(b) describes the pulse sequence to prepare and measure the phase of a superposition state.
Take nuclear spin for example, the qubit is defined in a rotating frame with frequency equalling to the energy splitting between $|0\uparrow\rangle$ and $|0\downarrow\rangle$ states. After polarized to $|0\uparrow\rangle$ by laser pulse, a resonant RF $\frac{\pi}{2}$ pulse brings the system to $\frac{1}{\sqrt{2}}(|0\uparrow\rangle+e^{i\varphi}|0\downarrow\rangle)$ state. The phase of this state is determined by the relative phase of the applied RF pulse, which is tunable in experiment. The phase of electron superposition state is prepared in the same way, with resonant MW pulses.}

The phase information of a superposition state is detected by converting it to population information
of the spin qubits and a following optical readout.
To eliminate the system error in long time measurement, we use a self-calibration measurement scheme
as shown in Fig.~2(a).
%\textcolor[rgb]{0.00,0.00,1.00}
{For each unknown state siting on the equatorial plane of the Bloch sphere,
 we measure the Rabi oscillations driven by two orthogonal microwave pulses (0~$^\circ$ and 90~$^\circ$),
 and compare the amplitudes of the two Rabi signal to extract the original phase information:
 $\varphi=\arctan\frac{A_x}{A_y}$, where $A_x$ and $A_y$ are the amplitudes of Rabi signals
  driven by 0~$^\circ$ and 90~$^\circ$ microwave pulses, respectively.}
  Note that this is a single spin experiment, and we need to repeat the pulse sequence many times to get a reliable signal to noise ratio (SNR).
  %For the convenience of description, let us name this $N$ repeat of pulse sequence as a 'single measurement',
%  which gives one output of phase.
{Figs.~2(c-e) presents the nuclear Rabi signal of 0.1~M, 0.4~M and 2~M repeat
of the pulse sequence in Fig.~2(b), with an input phase of 30~$^\circ$. The error bar of the data point represent standard deviation (SD) of 10 repeat measurements.}
   It is clear to see from Fig.~2(c) to Fig.~2(e) that the SNR is better as the measurement sequence is repeated more times.
  %Fig. 2 (f) summaries the dependence of measured phase and its standard deviation on repeat number,
 % the decrease of phase estimation error can be well described by central limit theorem.
 In Fig.~2(f), the decrease of phase estimation error can be well described by central limit theorem.

\begin{figure}
\begin{center}
\includegraphics[width=7.5cm]{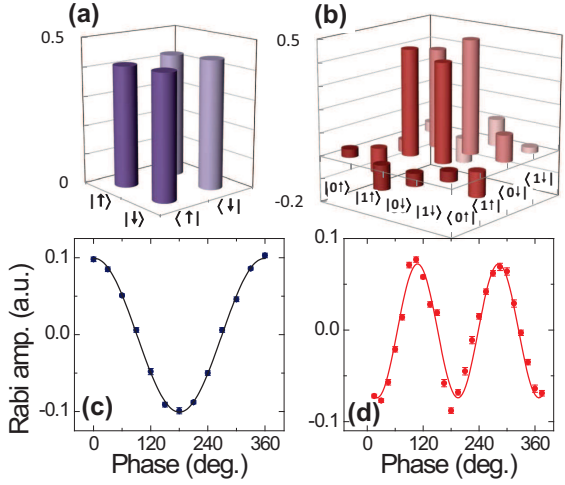}
\caption{\textbf{State tomography and phase relation. (a)} The state tomography result of {a
nuclear spin superposition state $\frac{1}{\sqrt{2}}(|\uparrow\rangle+|\downarrow\rangle)$}. \textbf{(b)} State tomography result of the
{electron-nuclear entangled state  $\frac{1}{\sqrt{2}}(|1\uparrow\rangle+|0\downarrow\rangle)$.} \textbf{(c)} Phase relation of an independent state.
\textbf{(d)} Phase relation of an entangled state.
{Compared with independent state, the phase relation of entangled state has double frequency dependence on input phase,  so a more precise phase estimation result can be achieved with entangled state.}
}
\end{center}
\label{Fig_3}
\end{figure}

To improve phase estimation accuracy, one can increase the repeat number, which means longer measurement time is needed.
An equivalent way is to employ more qubits. As mentioned before, the state of the multi-qubit system,
independent or entanglement, determines the accuracy limit of phase estimation.
For the investigated two-qubit system, the electron and nuclear spin can be prepared and measured independently.
Fig.~3(a) plots the state tomography result of a nuclear spin superposition state. Using such independent state
 (either nuclear spin or electron spin) will get a phase relation as depicted in Fig.~3(c), the amplitude of Rabi signal
 has cosine dependence on the phase of input state.

The electron and nuclear spin can be prepared in entangled state by combination of MW and RF pulses.
 {As shown in the upper pane of Fig.~2(b), after the first RF $\frac{\pi}{2}$ pulse (with phase $\varphi$) brings the system to  $\frac{1}{\sqrt{2}}(|0\uparrow\rangle+e^{i\varphi}|0\downarrow\rangle)$ state, a selective MW $\pi$ pulse of $|0\uparrow\rangle \Leftrightarrow|1\uparrow\rangle$ transition, which has relative phase $\varphi$ to the first RF pulse,
brings the system to $\frac{1}{\sqrt{2}}(|1\uparrow\rangle+e^{2i\varphi}|0\downarrow\rangle)$ state. The MW and RF channels are synchronized to the same clock reference and relative phase between them is calibrated before each measurement.}
Typical state tomography result of an electron-nuclear entangled state {($\varphi =0^\circ$)} is depicted in Fig.~3(b). It is worth noting that the dephasing time of electron spin {(0.7~$\mu$s, see Methods)} is very short compared with the typical manipulation time (e.g. $10~\mu$s for a flip operation) of nuclear spin, which limits the QIP applications of this entangled state \cite{GatewhileDD, DDGate, ESLAC09PRA}.
 However, in our phase estimation application, the sensitive phase information is converted to
  population right after its generation { and then only limited by $T_1$ of electron spin, which is about 5~ms. Meanwhile, the coherence of electron is less affected under microwave driving (see Methods), thus the phase of the entangled state is well preserved during the preparation and measurement.}  As shown in Fig.~3(d), the phase relation of the entangled state has double frequency dependence on the phase of input state, so the phase estimation using the entangled state of two qubit is more precise than that of using two state from independent single qubit.

\begin{figure}
\begin{center}
\includegraphics[width=8cm]{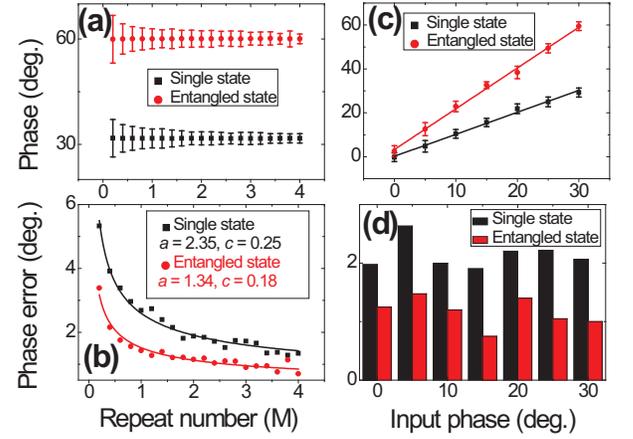}
\caption{\textbf{Dependence of phase error on repeat number and input phases.} \textbf{(a)}
Standard deviations of the entangled state and independent state against repeat number. Single
spin states (electron and nuclear) of the same input phase $\varphi=30^\circ$ are prepared and
measured independently, the phase extracted from each entangled state ``single measurement''
is $2\varphi=60^\circ$. \textbf{(b)} The phase uncertainty is fitted by function
$\delta \varphi =a/\sqrt {\nu }+c$ for both single spin state and the entangled state, where $\nu $ corresponds
to repeat number in unit of million (M). The curves show that the phase uncertainty,
phase error represented by $\delta \varphi $, of entanglement case
is apparently lower than the case of single spin state.
\textbf{(c)}  The phase estimation results of different input phases including
$0^\circ$, $5^\circ$, $10^\circ$, $15^\circ$, $20^\circ$, $25^\circ$ and $30^\circ$, the repeat
number is fixed to 1 M. \textbf{(d)} The phase error of different input phases, the repeat
number is fixed to 1 M.}
\end{center}
\label{Fig_4}
\end{figure}

\subsection{Entanglement-enhanced phase estimation}
%(SD$\delta\varphi$ vs N, SD$\delta\varphi$ vs $\varphi$)
To demonstrate the merits of entangled state over independent state in phase estimation application,
we compare their performances on different repeat number and different input phase, the measured
results are summarized in Fig.~4.
The experimental procedure is: firstly, single spin states (electron and nuclear)
of the same input phase ($30^\circ$) are prepared and measured independently.
Then the output phases extracted from the same repeat number are counted together,
no weight is added for either electron or nuclear spin states.
For a fair comparison with entangled state, half of the statistic samples ($\nu $) are extracted from electron spin states,
and the other half ($\nu $) are extracted from nuclear spin states.
In the case of entangled state phase estimation, the entangled states are prepared and measured
using the same repeat number {($\nu $)}.
%, and such `single measurement' are repeated equal times ($\nu $) to get comparable statistic results.
Note the same MW and RF channels are used to prepare the independent and entangled states.
%and a state of electron spin together with a state of nuclear spin are packaged to compare with one entanglement state of them.

As shown in Fig.~4(a) and (c), the phase extracted from each entangled state {measurement} is $2\varphi$,
 so the phase error of input phase is just half of the standard deviation from $\nu$ sample statistic (${\delta\varphi}/{2}$).
 For the independent-state input, the double sample number ($2\nu$) only
 suppresses phase error to ${\delta\varphi}/{\sqrt{2}}$ level, which is larger than entanglement-state input.

Explicitly, we would expect that the phase uncertainty $\delta\varphi $
proportion respectively to $1/\sqrt{\nu N}$ and $1/(\sqrt{\nu }N)$ for single spin state and the
entangled state with $N$ being 2 in our experiment since two-qubit entanglement
between electron spin and nuclear spin is used as the quantum resource.
In Fig.~4(b), we consider that identical measurement is repeated $\nu $ times,
then a general formulae, $\delta\varphi =a/\sqrt {\nu }+c$, is used to fit experimental data,
where $c$ is assumed to be a systematic error
depending on specific experimental setup.
The parameter $a$ for scheme of entangled state
should be smaller than that of the single spin scheme, corresponding to
smaller uncertainty about the phase, if we assume that the single spin state and the entangled state are
realized by the same physical state. Fig.~4(b) of the experimental data
demonstrates clearly that the precision of phase estimation is enhanced
by using entanglement which agrees well with theoretical expectation.
Here the discrepancy between theory and experiment is possibly due to
two related reasons: the electron spin state and the nuclear spin state are not the same, in particular
for their decoherence time, while their similarity is assumed theoretically;
the readout of NV centre system can only be by intensity of florescence
of the electron spin. {Besides the data processing method presented
here, we have also tried linear fitting in the log-log scale for SD as well as variance
(see Supplementary Note~2 and Supplementary Fig. 3 for detail).
The obtained results are in good agreement with the results in Fig.~4(b), which confirms the
validity of our conclusion.}

Fig.~4(c) and (d) show the phase estimation results of different input phases. %, the repeat number is fixed to $N=0.5$ M.
The phase error of entangled state is smaller than independent state in all input phases, which indicates the enhancement of phase
estimation accuracy by entanglement is phase independent.

\section*{Discussion}

As summarized in Fig.~4 by different figures of merit quantifying
the uncertainty of phase estimation, the entanglement-enhanced precision
is clearly shown by experimental data.
This experiment demonstrates the advantage of the quantum metrology scheme.
Practically by using quantum metrology, the measured physical quantity should have
the same interaction on the probe system no matter it is prepared as a single qubit
or entangled state. In our special designed experiment, the measured phases
are artificially encoded to the probe state such that the enhancement of
precision can be shown by entangled probe state.
However in principle, the confirmation of theoretical expectation by
experimental data provides a solid evidence that quantum phase estimation
is applicable in this solid state system.
%Comparing with other systems, e.g. photons, the data of each measurement in
%our system is useful for analysis, while joint measurement scheme in general should
%be performed in photon counting which may diminish the efficiency of the
%quantum metrology by photons.

{ In this experiment, we use repeating measurement to
overcome the low photon collection efficiency of NV center. The phase estimation accuracy
can be further improved by employing single-shot measurement technique, which is now
available in NV system \cite{SingleShotReadout,Hanson11Nature, SingleShot13PRL}. Although
the photon collection efficiency is not perfect ($<$ 20\%, not every measurement is stored and counted), the following two facts
guarantee the reliability of the demonstration: (1) we use the same scheme to measure
single and entangled states, that is, the phase information is finally converted to fluorescence
signal of NV center and detected. (2) The detection efficiency of the system is stable (though
not perfect as single-shot readout) for all the measurement, so we can directly compare the
measured phase noise of single and entangled states.}

{As the phase estimation accuracy is determined by the total
number of entangled qubits, a straightforward way to improve the phase accuracy is increasing
the involved spin number. The large amount of weakly coupled $^{13}$C nuclear spins around
NV center are one of the best candidates. With the assistance of dynamical decoupling on
center electron spin, up to 6 $^{13}$C nuclear spins can be coherent manipulated \cite{WeakC13ZhaoN, WeakC13Lukin, WeakC13Hanson}.
Multi-qubit application such as error correction has been demonstrated in this system
\cite{ErrorCorr2014Hanson, ErrorCorr2014WangY}.}

In conclusion, we report the first
room-temperature implementation of entanglement-enhanced phase estimation
in a solid-state system: the nitrogen-vacancy (NV) centre in pure diamond.
We demonstrate a super-resolving phase measurement with two entangled qubits
of different physical realizations: a NV centre electron spin and a proximal
${}^{13}$C nuclear spin. Thus, our results represent a more generalized and elemental
demonstration of enhancement of quantum metrology against classical procedure,
which fully exploits the quantum nature of the system and probes.

\section*{{Methods}}
\subsection{Cram\'{e}r-Rao bound and quantum Fisher information}
In the simplest version of the typical quantum parameter estimation problem,
we aim to recover the value of a unknown continuous parameter [say phase $\varphi$
in Fig.~1(a)] encoded in a fixed set of states $\rho_{\varphi}$ of a quantum system \cite{review}.
We can obtain a single result $\xi$ via performing a measurement on the system and it is
useful to express the measurement in terms of set of POVM $\{\hat{E}_{\xi}\}$. With
large number of measurements, it is possible to calculate the estimator
$\tilde{\varphi}(\xi)$ with the observation conditional probability density function (\emph{pdf}) of
result $\xi$ given the true values $\varphi$: $p(\xi|\varphi)=\textrm{Tr}(\hat{E}_{\xi}\rho_{\varphi})$.
When the number of measurements $\nu$ is sufficiently large and the estimation
is unbiased \cite{fan,fan2}, the root-mean square error for the statistical uncertainty
can be shown to obey the well-known Cram\'{e}r-Rao bound \cite{Cramer} given by
\begin{eqnarray}
\delta\varphi\equiv\sqrt{\sum_{\xi}[\tilde{\varphi}(\xi)-\varphi]^{2}p(\xi|\varphi)}\geq\frac{1}{\sqrt{\nu F(\varphi)}},\label{1}
\end{eqnarray}
where
$F(\varphi)\equiv\sum_{\xi}p(\xi|\varphi)\left[{\partial_{\varphi}\ln p(\xi|\varphi)}\right]^{2}$
is the Fisher information corresponding to the selected POVM and the conditional \emph{pdf}
of the result. Eq.~(\ref{1}) provides a lower bound for the achievable lower bound by choosing the
optimal measurement expressed by some POVM $\{\hat{E}_{\xi}^{opt}\}$ that maximizes the
Fisher information: $F_{Q}(\varphi)=\max_{\{\hat{E}_{\xi}^{opt}\}}F(\varphi)$
which is known as the quantum Fisher information (QFI) \cite{1994}.
%Via assigning the symmetric
%logarithmic derivative (SLD) $L$ through $2\partial_{\varphi}=L\rho_{\varphi}+\rho_{\varphi} L$,
%the QFI can be written as $F_{Q}(\varphi)=\textrm{Tr}(\rho_{\varphi}L^{2})$.
%The simplest but common case is when the state of the system is a pure state $\rho_{\varphi}=|\psi_{\varphi}\rangle\langle\psi_{\varphi}|$, and meanwhile one is
%able to write the QFI as \cite{escher}
%\begin{eqnarray}
%F_{Q}(\varphi)=4\left(\frac{\partial\langle\psi_{\varphi}|}{\partial\varphi}\frac{\partial|\psi_{\varphi}\rangle}{\partial\varphi}-\left|\frac{\partial\langle\psi_{\varphi}|}{\partial\varphi}|\psi_{\varphi}\rangle\right|^{2}\right).
%\end{eqnarray}
%For instance, consider that $|\psi_{\varphi}\rangle=\exp(-iH\varphi)|\psi\rangle$ with a
%Hermitian opertaor $H$ being the generator, $F_{Q}(\varphi)=4(\Delta H)^2$ and
%the Cram\'{e}r-Rao bound reduces to a simple form
%$\delta\varphi\geq{1}/({2\sqrt{\nu}\Delta H})$,
%given that $\Delta \hat{H}^{2}={\langle\psi|\hat{H}^{2}|\psi\rangle-\langle\psi|\hat{H}|\psi\rangle^{2}}$.

For the classical scheme with separable probe state $[(|0\rangle+e^{-i\varphi}|1\rangle)/\sqrt{2}]^{\otimes N}$, the lower
bound at best leads to the SQL $\delta\varphi_{se}\propto1/\sqrt{\nu N}$. To implement the quantum
counterpart of the Heisenberg limit $\delta\varphi_{en}\propto1/\sqrt{\nu}N$, we can choose the
GHZ state $(|0\rangle^{\otimes N}+e^{-iN\varphi}|1\rangle^{\otimes N})/\sqrt{2}$
as the optimal probe state. Consider that the qubit number $N=2$, the two-qubit  maximally
entangled state will obtain a $\sqrt{2}$ advantage against the separable state.

\subsection{Sample preparation}

High purity single crystal diamond (Element Six, N concentration $<$ 5~ppb) is used for this
experiment. There is almost no natural NV center in this diamond.  NV centres are produced
by electron implantation (7.5~Mev) and a following 2 hours vacuum annealing (at 800 $^{\circ}$C).
Due to the random distribution of $^{13}$C nuclear spins, the spin bath of individual NV
center can be very different \cite{Hyperfine}. We choose NV centers with nearby $^{13}$C
nuclear spins, which can be identified by the extra splitting in ODMR signal, to implement the
two-bit metrology scheme. Fig.~5(a) presents the physical structure of an NV center and a
nearby $^{13}$C nuclear spin. Fig.~5(c) is ODMR signal of this two-bit system. The coupling
strength between electron spin and the selected $^{13}$C nuclear is 12.8~MHz. Fig.~5(b)
shows two dimensional fluorescence image of the FIB-etched SIL. The cross-cursor marked
bright spot (blue) is the one used for this experiment.

\subsection{Coherence of electron spin and nearby nuclear spin}

In this pure diamond, the coherence of NV electron spin is dominated by the randomly
distributed $^{13}$C nuclear spins (natural abundance, 1.1\%). The dephasing time of
individual NV centres can be significantly different \cite{FID}, from less than 1~$\mu$s
to nearly 10~$\mu$s. From the free-induction decay signal of this NV center in Fig.~5(d),
we extract the dephasing time ($T_2^*$) of this electron spin, which is 0.72~$\mu$s and
much larger than the time consumption of single manipulation on it (about 70~ns, see Rabi
oscillation of electron spin in the same figure). It is worth noting that the dephasing time is
not the direct limitation of electron manipulation duration. The latter is usually named
$T_{1\rho}$ and can be characterized by the envelop decay time of electron spin Rabi oscillation \cite{DNP}.
The dephasing time of nuclear spin ($T_{2n}^*$= 270 $\mu$s) is much longer than that of electron spin.
Meanwhile, the used half $\pi$ pulse of nuclear spin is only several microseconds,
so the dephasing effect of nuclear spin can be ignored,
See Supplementary Fig. 1 for the FID signal of the nearby nuclear spin.

\begin{figure}
\begin{center}
 \includegraphics[width=7.8cm]{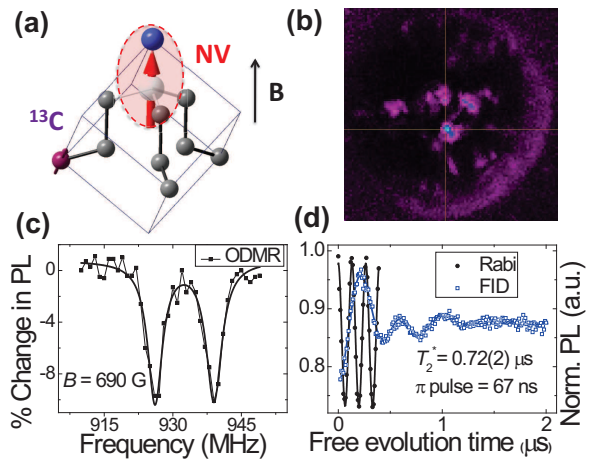}
\caption{ \textbf{Characterization of the two-qubit system.} \textbf{(a)} NV centre with a nearby $^{13}$C nuclear spin.
\textbf{(b)} 2D fluorescence scan of the FIB-etched solid immersion lens (SIL, 12~$\mu$m diameter). The bright spot is the investigated NV centre.
\textbf{(c)} ODMR of NV electron spin under magnetic field of 690~Gauss. The splitting is caused by the nearby $^{13}$C nuclear spin.
\textbf{(d)} Rabi oscillation and Free-induction decay (FID) of electron spin ($B=507$~Gauss). Due to the thermal fluctuation of the spin bath, centre electron spin picks up random phase during free precession and loses coherence. With the help of resonant MW pulses, electron spin can be flipped in short time, and is less affected by the bath noise.}
\end{center}
\label{Fig 5}
\end{figure}

\subsection{Coherent manipulation of electron spin and nuclear spin at ESLAC}

As mentioned in the main text, we work at the excited-state level anti-crossing (ESLAC) point to
achieve fast and high fidelity initialization of the electron-nuclear two-qubit system. Under an
external magnetic field of 507~Gauss (along the quantization axis of the selected NV) and laser
excitation (532~nm), the electron and nuclear spins are polarized simultaneously. Fig.~6(a) shows
ODMR spectrum of centre electron spin at such magnetic field. From the contrast difference of
two peaks, which correspond to $|\uparrow\rangle$ and $|\downarrow\rangle$ states of $^{13}$C
nuclear spin, we estimate the polarization rate of this nuclear spin is about 85\%
(in $|\uparrow\rangle$ state). Furthermore, by measuring the pulse-ODMR spectrum of electron
spin, we conclude that the host $^{14}$N nuclear spin is completely polarized under this magnetic field.

Fig.~6(b) shows the pulse sequence of electron and nuclear spin manipulation. After polarization with
high fidelity, both spin states can be manipulated with resonant MW (or RF) pulses. For electron spin,
the final state is readout by counting the fluorescence intensity of NV center, since $|0\rangle$ state
is brighter than $|1\rangle$ state. For nuclear spin state, we use a mapping gate, which is composed
by a weak  pulse of $|0\uparrow\rangle \Leftrightarrow|1\uparrow\rangle$ transition, to transfer
the its state to electron spin and then readout optically. For example, an unknown nuclear spin state
of $|0\rangle\otimes(\alpha|\uparrow\rangle +\beta|\downarrow\rangle)$ is transferred to
$\alpha|1\uparrow\rangle +\beta|0\downarrow\rangle$ after applying the mapping gate.  We
carefully tuned the microwave power and pulse duration to maximum the flip efficiency while avoiding
the unwanted non-resonant excitation. By comparing the Rabi amplitude of nuclear spin (Fig.~6(d), with mapping gate)
 and electron spin (Fig.~5(d), without mapping gate),
 we conclude that the mapping gate has transfer efficiency of more than 92\%.

Figs.~6(c) and 6(d) present the pulse-ODMR spectrum and Rabi oscillation of the nearby nuclear spin when
electron spin is at $|0\rangle$ state. The resonant frequency of this nuclear spin is 495~kHz, which is
smaller than the Larmor frequency of $^{13}$C nuclear spin under this magnetic field (542~kHz). We
attribute this modification to the ``enhance effect'' of center electron spin. As the nuclear spin is close
to the electron spin, the nonsecular terms of their dipole interaction contribute some electronic
character to the nuclear-spin levels and modify its magnetic moment \cite{Childress06Science}. The
Rabi frequency of nuclear spin is about 100~kHz, which reaches 20\% of the Zeeman splitting, such fast
manipulation also benefits from the electron enhance effect. We discuss the validity of rotating wave
approximation (RWA) in Supplementary Note 1.

\begin{figure}
\begin{center}
 \includegraphics[width=7.8cm]{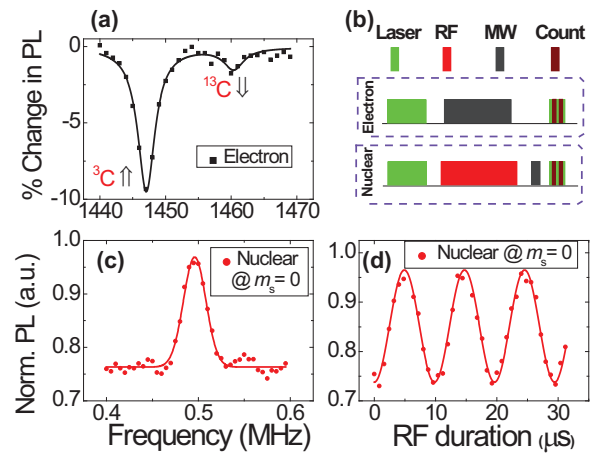}
\caption{\textbf{Coherent manipulation of electron spin and nuclear spin at ESLAC}.  \textbf{(a)} ODMR spectrum of electron spin at $B=507$ Gauss. \textbf{(b)} Pulse sequence to manipulate electron and nuclear spins at ESLAC. \textbf{(c)} ODMR spectrum and \textbf{(d)} Rabi oscillation of the nearby $^{13}$C  nuclear spin. At ESLAC, electron spin and nearby nuclear spins (including host $^{14}$N  nuclear spin and nearby $^{13}$C  nuclear spins) can be polarized by a short laser pulse. Then spin states are manipulated by resonant MW or RF pulses. Electron spin states are readout by counting the fluorescence intensity of NV centre; nuclear spin states are mapped to electron spin and readout in the same way. The resonant frequency of this nuclear spin is 495 kHz, which is slightly modified by centre electron spin.}
\end{center}
\label{Fig 6}
\end{figure}

\subsection{Synchronization of pulse generators and phase calibration}
Synchronization of the MW and RF generators is one of the main challenges in this experiment. We use the same clock reference for all the generators. For each cable connection and pulse sequence, we measure the phase of prepared state as we scan the phase of input MW pulses. This gives us a ¡°phase relation¡± between MW and RF channels, which is used to compensate the difference between the two rotating frames. We check the ¡°phase relation¡± before and after each data acquisition. The phase drift of our system is about 2 $^\circ$ in 2 hours¡¯ measurement.

\subsection{Data normalization and state tomography}

Since the population information of electron spin is the only directly measurable signal in
NV system, we normalize all the data to the fluorescence intensity of electron spin $|0\rangle$
state. Specifically, we apply two readout pulses (300~ns) at the end of each measurement.
See pulse sequences in Fig.~2(b) and Fig.~6(b). The first readout pulse gets the instant
population information of NV electron spin, and the second readout pulse (1~$\mu$s later)
records a reference for the first one, as electron spin is polarized to $|0\rangle$ again after
the 1~$\mu$s laser excitation. The ratio between the first signal and the reference signal is
used for further data analysis, such as phase estimation or state tomography.

To carry out state tomography, we adopt the method detailed in Ref.~\cite{Wrachtrup08entangle,DDGate}.
Total three working transitions, $|0\uparrow\rangle \Leftrightarrow|0\uparrow\rangle$,
$|0\uparrow\rangle \Leftrightarrow|1\uparrow\rangle$ and $|0\downarrow\rangle \Leftrightarrow|1\downarrow\rangle$
are selected. The real and imaginary parts of the matrix elements in each working transition
are measured by using RF (or MW) pulses of 0$^{\circ}$ and 90$^{\circ}$ phases, respectively.
Other three transitions are measured in the same way, but extra transfer pulses are added
before Rabi measurement in the working transitions.
The full procedure of state tomography can be found in Supplementary information (Supplementary Fig. 2).

\textbf{Acknowledgement}
This work was supported by the
National Basic Research Program of China (``973'' Program under Grant Nos. 2014CB921402 and 2015CB921103),
National Natural Science Foundation of China (under Grant No. 11175248),
and
the Strategic Priority Research Program of the Chinese Academy of Sciences (under Grant Nos. XDB07010300 and XDB01010000).

\textbf{Author Contributions}
X.-Y.P. and H.F. designed the experiment. X.-Y.P. is in charge of
experiment, H.F. is in charge of theory. G.-Q.L., Y.-C. C.
and X.-Y.P. performed the experiment. Y.-R. Z. and J.-D. Y
carried out the theoretical study. G.-Q.L. and Y.-R. Z. wrote the paper.
All authors analysed the data and commented on the manuscript.

\textbf{Competing Interests}
The authors declare that they have no competing financial interests.

\textbf{Correspondence}
Correspondence and requests for materials should be addressed to
X.-Y.P or H.F.

\renewcommand{\thefigure}{S\arabic{figure}}
% redefine the command that creates the equation no.
\setcounter{figure}{0}  % reset counter
\appendix
\section{Supplementary Materials}
\begin{figure}[b]
\begin{center}
\includegraphics[width=8.5cm]{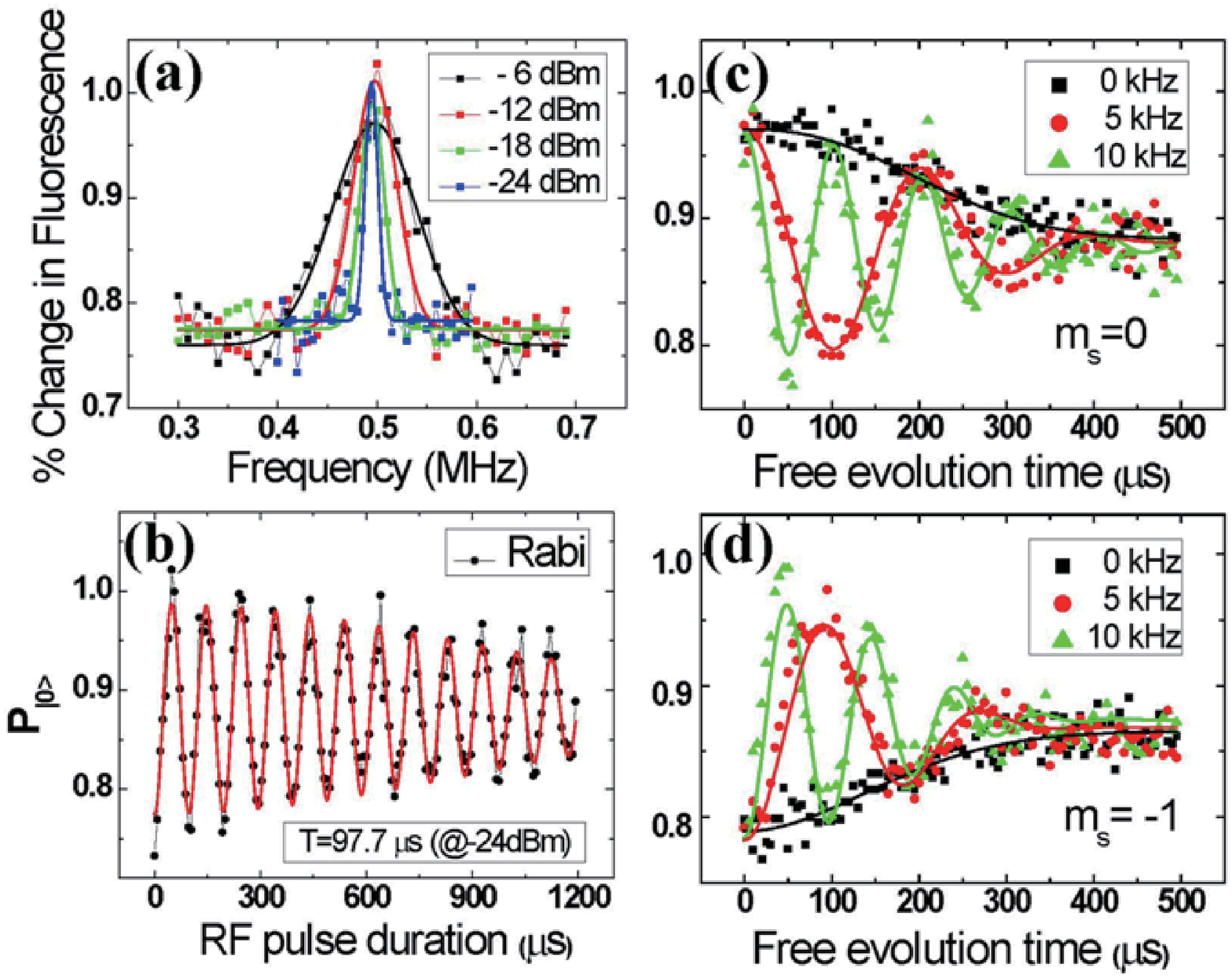}
\caption{\textbf{Coherence of nuclear spin.} \textbf{(a)} Pulse-ODMR of ${}^{13}$C nuclear spin with RF
pulses of different driven power. The resonant frequencies are the same for all the measurement
(evidence of weak driven). \textbf{(b)} Rabi oscillation of nuclear spin under weak RF driven. The measured Rabi
frequency is about 10~kHz, with typical envelop decay time of $T_{1\rho}= 1.3$~ms. \textbf{(c)} FID of nuclear
spin when electron spin is at $m_S=0$ state, with dephasing time of $T_{2n}^*(m_{s}=0)=270$~¦Ìs. \textbf{(d)}
FID of nuclear spin when electron spin is at $m_S=-1$ state, with dephasing time of $T_{2n}^*(m_{s}=-1)=212$~¦Ìs.
The square, circle and triangle are experiment data with detuning of 0~kHz, 5~kHz and 10~kHz, respectively.
Solid lines are fitting to them.}
\end{center}
%\label{Fig_2}
\end{figure}
\subsection{Supplementary Note 1: Validity of rotating wave approximation and coherence of nuclear spin}
As mentioned in the main text, the achieved Rabi frequency of nuclear spin is not small compared 
with the energy gap ($|0\uparrow\rangle\rightarrow|0\downarrow\rangle$), thus we need to 
evaluate whether rotating wave approximation (RWA) still works well under this circumstance. 
We measure nuclear pulse-ODMR spectrum and Rabi oscillation with RF pulses of different driven 
powers, as shown in Supplementary Fig.~1(a-b). The minimum Rabi frequency is only $10$~kHz with 
a RF power of $-24$~dBm (at signal generator), which is much less than the energy gap of nuclear 
spin and RWA works well at this power. We then choose $\pi$ pulse of this RF power ($49$~$\mu$s) 
to measure pulse-ODMR spectrum of nuclear spin. The resonant frequency between $|0\uparrow\rangle$ 
and $|0\downarrow\rangle$ transition is $495$~kHz under this weak driven power. We find that the 
resonant frequencies are the same for all the measured RF pulses, including the one, which 
corresponds to a Rabi frequency of about $20\%$ of nuclear spin energy gap. Therefore, we conclude that 
RWA works well for all the measurements in this experiment. This conclusion conforms with the results in 
Supplementary Ref.~1 and Supplementary Ref.~2, where RWA works well with a Rabi frequency less than 
half of the energy gap of two-level system (NV electron spin).

We then consider the coherence of nuclear spin. Supplementary Fig.~1(c) and (d) present the FID 
signals of nuclear spin under ESLAC, for both $m_S=0$ and $m_S=-1$ states of electron spin. The 
dephasing time of nuclear spin ($T_{2n}^{*}=270$~$\mu$s for $m_S=0$ state and 210~$\mu$s for $m_S=-1$ state) is shorter 
than the result in Supplementary Ref.~3 This may be caused by the complicated spin bath of this NV 
center. However, similar to the case of electron spin, the dephasing time is not the limitation of 
nuclear manipulation duration. The Rabi envelope decay time of this nuclear spin ($T_{1\rho}$) is more 
than $1$~ms. The half $\pi$ pulse, which is used to generate the superposition state of nuclear spin, is 
only $5$~$\mu$s and much shorter than the dephasing time. So we ignore the dephasing effect of nuclear 
spin in the metrology experiment.

\begin{figure}[b]
\begin{center}
\includegraphics[width=8.5cm]{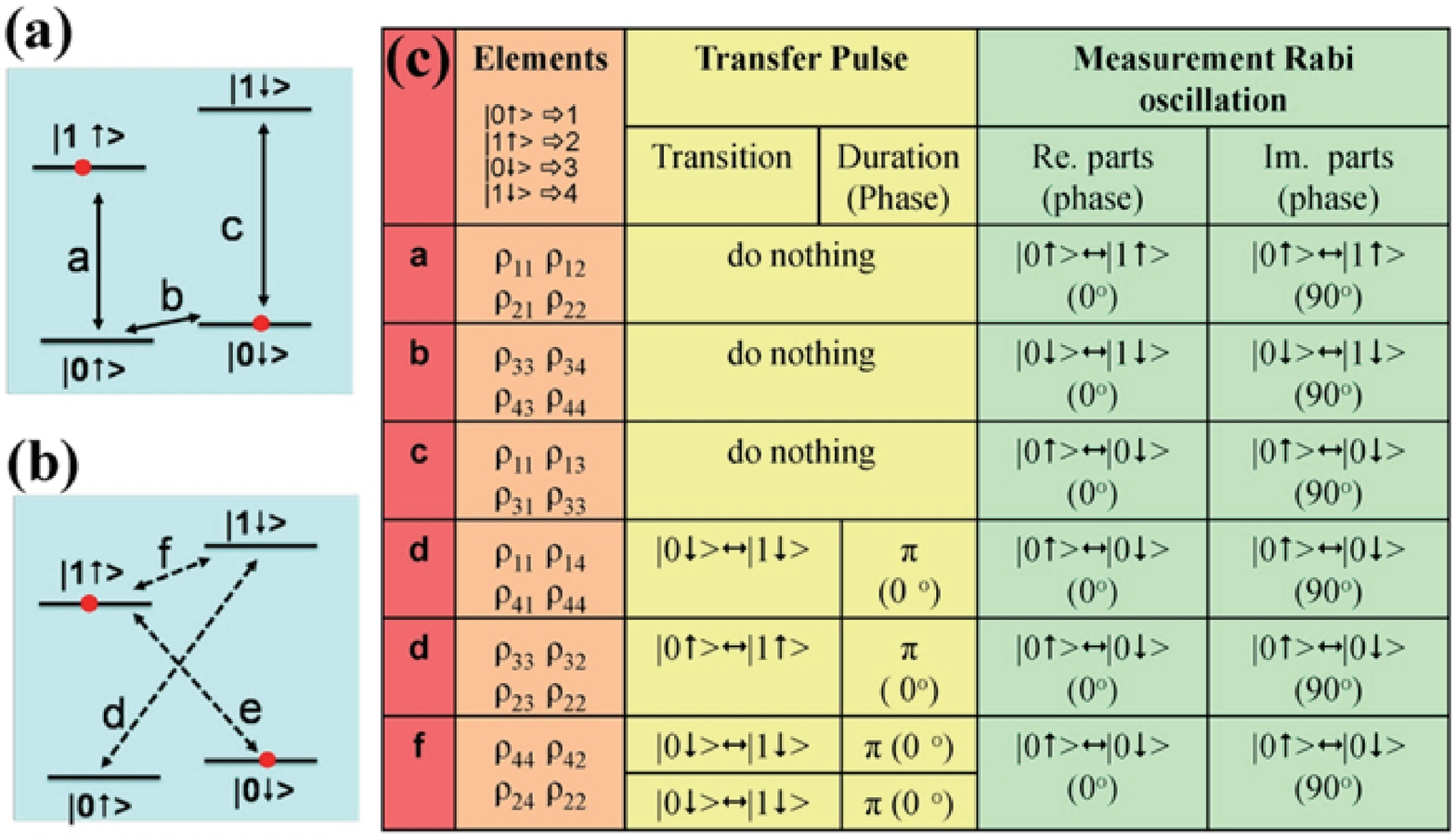}
\caption{\textbf{Procedure to carry out state tomography on electron-nuclear two-qubit system.} \textbf{(a)} 
The three solid arrows are selected working transitions, which can be driven by MW/RF pulse directly. 
The real and imaginary parts of the matrix elements in each working transition are measured by RF (or MW) 
pulses of $0^{\circ}$ and $90^{\circ}$ phases, respectively. \textbf{(b)} For the other three transitions (dash arrow), 
one or two pulses are applied to transfer the state information to working transition 
($|0\uparrow\rangle\rightarrow|0\downarrow\rangle$) before Rabi measurement. \textbf{(c)} Pulse duration 
and phase of state tomography. The diagonal elements of density matrix are measured three times, and the 
mean of these measurements are used.}
\end{center}
%\label{Fig_2}
\end{figure}

\begin{figure}[t]
\begin{center}
\includegraphics[width=8.5cm]{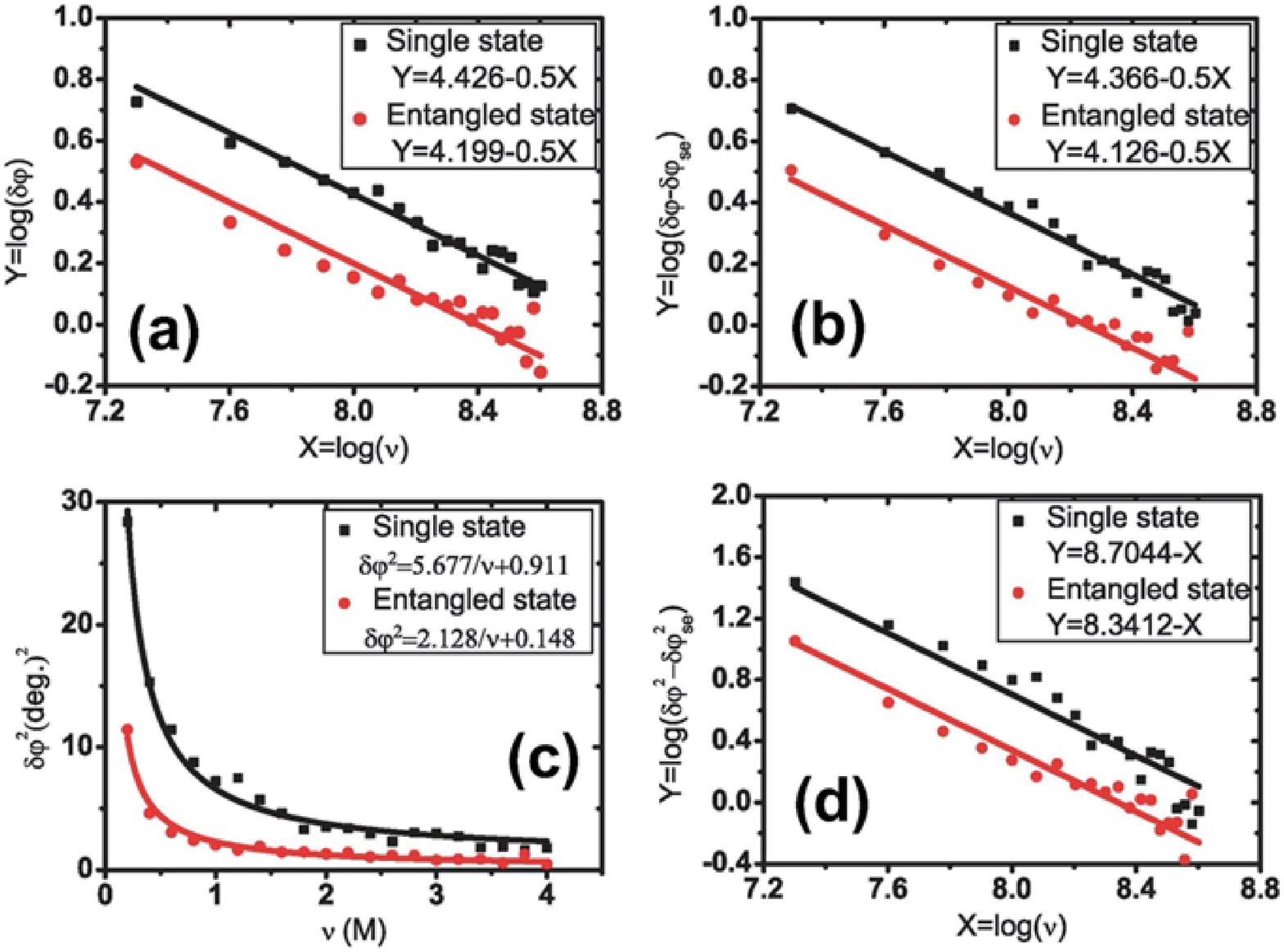}
\caption{\textbf{Data processing for standard deviation and variance of phase.} \textbf{(a)} Linear
fitting for standard deviation $\delta\varphi^2$ against repeat number $\nu$ in the log-log scale
with adjusted R-squares 0.967 and 0.881 for single state and entangled state. (b) Linear fitting for
standard deviation subtracted the system error $\delta\varphi_{\textrm{se}}$ in the log-log scale
for single state and entangled state. Adjusted R-squares are 0.962 and 0.904. (c) The variance of
phase is fitted by function $\delta\varphi^2=a/\nu+c$ where c represents the squared system error
$\delta\varphi^2$. (d) Linear fitting for variance subtracted the squared system error in the log-log
scale. Adjusted R-squares are 0.916 and 0.893 for single state and entangled state, respectively.}
\end{center}
%\label{Fig_2}
\end{figure}

\subsection{Supplementary Note 2: Other data processing methods for standard deviation and variance}
Regardless of system error, the variance of phase $\delta\varphi^2$ with a sufficiently large number of 
measurements $\nu$ will be approximately normally distributed as $\delta\varphi^2\propto1/\nu$, 
which is based on the classical central limit theorem. This fact can also be explained by the additive 
property of Fisher information and is shown in Eq.~(1) in the main text. Therefore, in Fig.~4(b) we set 
the exponent of $\nu$ as $0.5$ for standard deviation (SD) and a conclusive result is given. In 
Supplementary Fig.~3, we try other data processing methods for the presentation of the 
entanglement-enhanced metrology. In Supplementary Fig.~3(a), we use the linear fitting in the 
log-log scale to analyze the SD, $\delta\varphi^2$, against $\nu$. Ideally the slope should be 0.5 and the 
intercept gives the value that represents the enhancement. However, there always exists the system 
error which will insult the linear analysis and give an inconclusive result for large number of 
measurement $\nu\sim1$~M. We thus fix the slopes as 0.5 and give the enhancement by reading 
the intercepts in Supplementary Fig. 3(a). The adjusted R-square, $\bar{R}^2$, (ranging from 0 to 1 
with larger number indicating better fitting) for single state and entangled state are 0.967 and 0.881, 
respectively. In Supplementary Fig.~3(b), the system error, $\delta\varphi_{\textrm{se}}$, read in 
Supplementary Fig.~3(c) is taken into consideration and subtracted out before analyzing, which leads to 
better adjusted R-squares: 0.962 and 0.904 for single state and entangled state, respectively. Therefore, 
we conclude that the data processing method we use in the main text for SD provides a better fitting to 
the experimental data. We also use the function  $\delta\varphi^2=a/\nu+c$ to fit experimental data of 
variance, see results in Supplementary Fig.~3(c). Taking out the effect of squared system error 
$\delta\varphi^2_{\textrm{se}}$, we use the linear fitting in the log-log scale and present the enhancement 
in Supplementary Fig.~3(d) with $\bar{R}^2$ being 0.916 and 0.893 for single state and entangled 
state, respectively. It shows that the data processing method we use for SD in the main text is 
better than that for variance.

\clearpage

\end{document}